\documentclass[letterpaper,twoside,12pt]{article}
\usepackage{amssymb}
\usepackage{amsmath}
\usepackage{latexsym}
\usepackage{verbatim}
\usepackage{graphicx}
\usepackage{epsfig}
\usepackage{pstricks}
\usepackage{stmaryrd}
\usepackage{rotating,graphicx}
\usepackage[english]{babel}
\usepackage{setspace}
\usepackage[english]{babel}
\usepackage[utf8]{inputenc}
\usepackage[T1]{fontenc}
\usepackage{lmodern}

\begin{document}

\newcommand{\bra}[1]{\langle #1|}
\newcommand{\ket}[1]{|#1\rangle}
\newcommand{\braket}[2]{\langle #1|#2\rangle}

\begin{Large}
\begin{center}
\textbf{The Black Hole Information Paradox and the Collapse of the Wave Function}\\
\end{center}
\end{Large}

\begin{center}
\begin{large}
Elias Okon\\
\end{large}
\textit{Instituto de Investigaciones Filos\'oficas, Universidad Nacional Aut\'onoma de M\'exico, Mexico City, Mexico.\\ E-mail:} \texttt{eokon@filosoficas.unam.mx}\\[.5cm]
\begin{large}
Daniel Sudarsky\\
\end{large}
\textit{Instituto de Ciencias Nucleares, Universidad Nacional Aut\'onoma de M\'exico, Mexico City, Mexico.\\ E-mail:} \texttt{sudarsky@nucleares.unam.mx}\\[.5cm]
\end{center}

\noindent 
The black hole information paradox arises from an apparent conflict between the Hawking black hole radiation and the fact that time evolution in quantum mechanics is unitary. The trouble is that while the former suggests that information of a system falling into a black hole disappears, the latter implies that information must be conserved. In this work we discuss the current divergence in views regarding the paradox, we evaluate the role that objective collapse theories could play in its resolution and we propose a link between spontaneous collapse events and microscopic virtual black holes.\\

\onehalfspacing
\section{Introduction}
The notion of measurement is essential to the standard formulation of quantum mechanics. Therefore, such formalism necessarily divides the world into two parts: the observer and what is observed. But, what distinguishes certain subsystems to play the role of observers?, or some interactions to play the role of measurements? The formalism offers no clear rules on the matter, and the problem is that its predictions crucially depend on the issue. This, in essence, is the \emph{measurement problem} of quantum theory. Much has already been said about the problem and we will not dwell much on it except to highlight one of the most promising paths towards its resolution, namely, the \emph{dynamical reduction} or \emph{objective collapse} models. Such theories modify the quantum evolution laws with the addition of non-unitary terms designed to overcome the problem.

In this work, we will connect those proposals with another open issue that is currently drawing the attention of an important sector of the theoretical physics community: the black hole information loss paradox. The issue arises because black holes appear to lose information but such conclusion goes against the fact that unitary quantum evolution preserves it. We will show, however, that objective collapse models contain enough resources to address the issue. To do so, we will first discuss the information loss paradox  in section \ref{P}  and in section \ref{D} we will describe the role that objective collapse models could play in its resolution. Next, in section \ref{B}  we will propose a link between spontaneous collapse events and  virtual black holes and finally,  in section \ref{C}, we will present our conclusions.
\section{The information loss paradox}
\label{P}
General relativity establishes that the end point of evolution for sufficiently massive stars are black holes. Moreover, even if at intermediate times the evolution of such systems involves complex dynamics, they eventually settle down into a stationary state that is described, at least in the exterior region, by one of the stationary black hole solutions admitted by the theory. Such solutions turn out to be completely characterized by a very small number of parameters, namely, the mass, charge and angular momentum appearing in the Kerr-Newman black holes. In fact, a large set of results, collectively know as \emph{black hole uniqueness theorems}, ensures that as long as one considers just the long range fields known to exists in nature, i.e., the gravitational and the electromagnetic fields, then these solutions represent the complete class of stationary black holes. What is more, even if one wishes to consider some other hypothetical fields, the so-called \emph{no-hair theorems} indicate that, over a large class of theories, the set of stationary solutions is not enlarged. 

All these results indicate that when a body collapses to form a black hole, the large amount of information corresponding to the full characterization of the collapsing body (type of matter, multipole moments of the initial mass distribution, etc.) is simply lost. Note however that this loss of information refers only to that which is, in principle, available to the observers in the exterior region. That is because, in principle, the whole space-time, as well as the complete state of the matter fields, might be recovered using data located both in the outside and the inside of the black hole.\footnote{Technically, this is reflected in the fact that, while at earlier times we can find Cauchy hyper-surfaces completely contained in the outside region, at very late time we can only find Cauchy hyper-surfaces that have parts in the exterior and parts in the interior regions.} Therefore, this loss of information is not really puzzling: all that happens is an emergence of a region of ``no escape,'' and a shift of some of the information from the outside into such region.
 
When quantum theory is brought to bear in this situation, the problem acquires novel features. That is because, as was shown in \cite{Haw:75}, quantum mechanical effects cause black holes to radiate and lose mass; and, unless something strange happens (like the formation, perhaps due to quantum gravity effects, of a stable remnant), the process is supposed to go on until the black hole completely disappears. But, how do these quantum mechanical effects modify the loss of information issue? The truth is that there is no consensus on the issue since it depends not only on what one assumes about the singularity within the black hole, but also on what one thinks about the nature of space-time and even on what one believes our physical theories should be about.
 
Let us start with the singularity. General relativity indicates that, under quite general conditions, the formation of black holes also leads to the development of singularities. In fact, the results know as \emph{singularity theorems} are, if anything, more stringent than those governing the formation of black holes. These singularity results indicate that, as long as the matter that is undergoing gravitational collapse satisfies some very mild and reasonable energy conditions (such as always possessing positive densities), then the formation of singularities is an inescapable result of the theory. Additionally, the formation of black holes is tied to a very interesting conjecture, supported by a relatively important body of evidence, \cite{CCC}. This so-called \emph{cosmic censorship conjecture} indicates that, for the observers at infinity, the singularities generated as the result of gravitational collapse are always hidden behind an event horizon.

A singularity can be thought of as a breakdown in the geometrical structure of space-time. Therefore, it represents a ``place'' in which the basic postulates of general relativity, involving a smooth space-time manifold, simply cease to hold. Singularities, then, seem to imply the collapse of our best classical theory of space-time. A possible way out of this conclusion is to consider the singularity (or, more precisely, a region arbitrarily close to it) as the boundary of space-time. If that is the case, the information loss paradox could be addressed by holding that information ``ends up'' at, or, if one wants to use a more pictorial language, ``escapes through,'' the singularity. If so, one would take Figure 1 as depicting all that can be said by the theory.
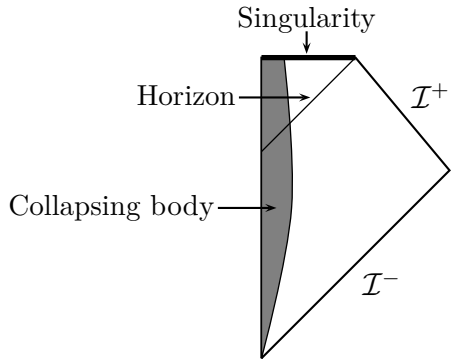
\begin{figure}[h]
\centering
 \begin{pspicture}(4.5,5)
 \psline(1,0)(1,4)(2.25,4)(3.5,2.5)(1,0)
\pscustom[linewidth=.5pt,fillstyle=solid,fillcolor=gray]{
\pscurve(1,0)(1.4,2)(1.3,4)
\psline[liftpen=1](1,4)(1,0)}
\psline[linewidth=2pt](1,4)(2.25,4)
\psline[linewidth=.5pt](1,2.75)(2.25,4)
\rput(2.6,1){$\mathcal{I}^-$}
\rput(3.25,3.5){$\mathcal{I}^+$}
\rput(0,3.5){\small Horizon}
\psline{->}(.7,3.47)(1.65,3.47)
\rput(1.6,4.5){\small Singularity}
\psline{->}(1.6,4.35)(1.6,4.05)
\rput(-1,2){\small Collapsing body}
\psline{->}(.42,2)(1.2,2)
 \end{pspicture}
\caption{Penrose (i.e., conformally compactified, with null lines at 45º) diagram for a collapsing spherical body. $\mathcal{I}^+$ and $\mathcal{I}^-$ denote past and future null infinity.}
\end{figure}

In any case, it is widely believed that a theory of quantum gravity will be able to cure these singularities. The idea is that such a theory will be able to describe, perhaps in a language that is much more general than the space-time language appropriate to the classical theory, the part of the system that would have corresponded to the singularity in the classical setting. Of course, at this point not much can be said about what will substitute the singularity in a fully quantum version of the theory. However, given that: i) at the classical level, the singularity is well within the event horizon, and ii) the regions where quantum gravity can be expected to produce strong deviations from the classical picture are expected to be ``close'' to the singularity, it is generally assumed that whatever quantum gravity does, it will not dramatically alter the classical picture in the region accessible to the outside observers. As a result, the only trace of quantum gravity that is generally given a non vanishing likelihood of outlasting the evaporation process is something like a stable Planck mass remnant that, in principle, would be accessible to the outside observers after they have witnessed the complete evaporation of the black hole. However, such remnant is not considered capable of altering, in a substantial manner, the picture regarding the information issue. That is because its information content would be bounded by the number of its internal degrees of freedom and such number is not expected to be large given the small size and small energy that the remnant would have relative to the initial mass of the collapsing body (the difference being of 38 orders of magnitude for a solar mass black hole). Of course, this expectation is an extrapolation of a pattern observed elsewhere in nature, but in the absence of a theory of quantum gravity we can never be sure that a dramatic departure of said pattern is not going to occur in this case. At any rate, and as is done in most of the discussions on this topic, we will not contemplate this exotic option any further.
 
Quantum gravity effects confined to the close vicinity of the ``would be singularity'' are, then, generally not believed to play an important role regarding the amount of information to be recovered from the objects that lead to the formation of the black hole. On the other hand, the Hawking radiation is expected, on very general grounds, to dramatically alter the picture shown in Figure 1. The idea is that the radiation that will reach very distant observers will be carrying positive energy. As a result, on energy conservation grounds, one expects the mass of the black hole, as seen by those observers, to decrease. In fact, space-time is expected to ``respond'' to the energy and momentum of this Hawking emission according to something like a semiclassical Einstein's equation. That is, the quantum energy and momentum of the radiation are expected to act on the curvature of space-time in the same way that classical energy and momentum do. Therefore, Figure 1 should be replaced by a picture where distant observers see the black hole as having a diminishing mass that (ignoring the possibility of a quantum gravity remnant) will go to zero in a finite time. In such case, space-time, after the evaporation is complete, should become the trivial Minkowski space-time (see Figure 2).
 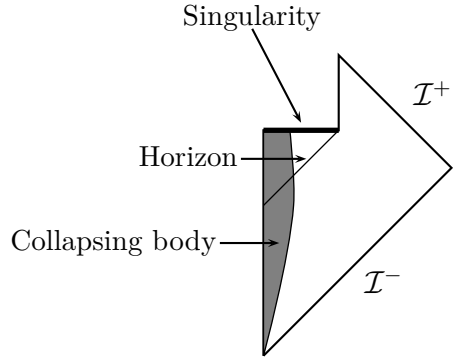
\begin{figure}[h]
\centering
 \begin{pspicture}(4.5,5)
 \psline(1,0)(1,3)(2,3)(2,4)(3.5,2.5)(1,0)
\pscustom[linewidth=.5pt,fillstyle=solid,fillcolor=gray]{
\pscurve(1,0)(1.4,2)(1.35,3)
\psline[liftpen=1](1,3)(1,0)}
\psline[linewidth=2pt](1,3)(2,3)
\psline[linewidth=.5pt](1,2)(2,3)
\rput(2.6,1){$\mathcal{I}^-$}
\rput(3.25,3.5){$\mathcal{I}^+$}
\rput(0,2.66){\small Horizon}
\psline{->}(.7,2.62)(1.57,2.62)
\rput(.85,4.5){\small Singularity}
\psline{->}(.85,4.35)(1.5,3.1)
\rput(-1,1.5){\small Collapsing body}
\psline{->}(.42,1.5)(1.21,1.5)
 \end{pspicture}
\caption{Penrose diagram for a collapsing spherical body taking into account Hawking's radiation.}
\end{figure}

Now, we can consider the loss of information issue in the following terms. We start with a particular state of the quantum matter fields characterized by some initial pure quantum state. Space-time, in turn, corresponds to a situation where, according to distant observers, at early times there is a matter-energy inflow that generates a black hole, while at late times there is a compensating outflow of energy which leaves an essentially empty space-time. Of course, the more accurate picture that emerges from the detailed general relativistic and quantum field theoretic analysis involves a singularity. This singularity, as we said, indicates that we are dealing with a regime where our theories break down and so they are no longer reliable regarding predictions. In the classical situation, this is remedied if the cosmic censorship conjecture is valid. In that case, far away observers would simply conclude that, regarding regions of the universe that are accessible to them, there is no loss of predictably. However, the inclusion of quantum effects leads to a picture where, at late times, there is no residual evidence of regions that are inaccessible to anybody (beyond the fact that the past is inaccessible to everybody) and so it seems that a problem remains. 

Indeed, even if one takes into account Hawking's radiation, one could still point out that the singularity is part of the full space-time, and that ignoring it would lead to an apparent breakdown of predictability for asymptotic observers describing ``late time'' physics. This seems to be the view of part of the community that argues that there is no paradox in the fact that the evaporation of a black hole is associated with an apparent information loss, reflected at the quantum level by the fact that an initially pure state evolves, as far as exterior observers are concerned, into a mixed state. The idea is that, if we want to deal with regions where our theories are valid, we need to include an extra boundary of space-time that separates the singularity from the regions that can be properly considered as a ``space-time.'' Such additional boundary could then be naturally associated with ``data'' regarding physical fields that, together with the data contained in the region accessible to distant observers, would comprise all the information that was present initially. At the quantum mechanical level, one would have to associate a quantum state to this part of the boundary, and such state, together with the state characterizing the fields outside and the correlations between them, would certainly be a pure quantum state.

The point of view expressed in the previous paragraph, however, does not seem to be satisfactory for many other researchers, specially those who work on approaches that attempt to produce a workable and self-consistent theory of quantum gravity. That is because, as we said, in such community it is generally believed that quantum gravity will resolve the singularity. As a result, the inclusion of an extra boundary is seen as something completely \emph{ad hoc} and uncalled for, and even worse, as something that removes from consideration the very regime for which quantum gravity theories are devised. Another generalized assumption of the quantum gravity community has to do with the fact that a theory of quantum gravity must incorporate all the general aspects of standard quantum theory. Therefore, if according to quantum theory, time evolution is always unitary --implying, among other things, that information must be preserved in the sense that, given a state at any time one can always predict or retrodict the state at any other time--, then that must also be the case for quantum gravity. 

The above, however, leads to a puzzle: if quantum gravity removes the singularity (and thus the need to incorporate an extra boundary of space-time) and quantum gravity does not account for any violation of unitarity or non-conservation of information, then even in the case of a black hole formation and evaporation, the quantum state at late times should be unitarily related to the quantum state at early times. The problem is that reconciling this with the picture that we saw emerges from considerations based on general relativity and quantum field theory, in regimes that the two theories ought to be valid, has proven to be extremely difficult. 

Of course, if this reconciliation turns out to be impossible, we would have a paradox. In fact, recent work on the, so-called, \emph{firewalls},\footnote{See \cite{Fire:12}; also see \cite{Br1,Br2} for a similar prediction from different assumptions.} seems to point in this direction. The idea is that a black hole evaporation process seems to imply that both unitarity and the equivalence principle cannot be true at the same time. This is because, on one side, for Hawking's radiation to occur, the emitted particles must get entangled with the ``twins'' that fall into black hole. On the other, if information is to come out with the radiation, then each emitted particle must also get entangled with all the radiation emitted before it. However, the \emph{monogamy of entanglement} holds that a quantum system cannot be fully entangled with two independent systems at the same time an so, unitarity and the equivalence principle cannot coexist. In \cite{Fire:12} it is suggested that we must forego the equivalence principle, allowing the event horizon to become a firewall. We, however, find it much wiser to do without unitarity. After all, the predictions of quantum mechanics are consistent with what we in fact perceive only after unitarity is broken. What we propose, then, is to study the black hole formation and evaporation process from the point of view of a quantum theory which incorporates, at the fundamental level, some kind of non-unitary evolution. A theory which allows for information to be lost but not only in exotic scenarios such as black holes but also, albeit in a smaller degree, in all situations and at all times. Of course, theories with such characteristics already exist in the form of objective collapse or dynamical reduction models (see \cite{SEP, Pea:07} for a general overview). The motivation behind such theories is to construct an alternative quantum formalism which solves the measurement problem. In order to do so, they modify the dynamical equation of the standard theory, with the addition of stochastic and nonlinear terms, such that the resulting theory is able to deal both with microscopic and macroscopic systems. In the next section we describe how this type of models may help in the solution of the information loss paradox.
\section{Dynamical reduction and breakdown of unitarity}
\label{D}
From the above discussion it should be clear that the analysis of the black hole evaporation process changes dramatically once one accepts, at the fundamental quantum level, a departure from unitarity. If unitarity is universally broken, and information is generically lost, then the fact that black holes lose information stops being that surprising and problematic. Therefore, in principle, such modification removes the impediment for black hole physics, including their eventual evaporation, to be described at the fundamental level using the same laws that we use for any other physical system. The interesting question, though, is whether it is possible to solve the information loss paradox quantitatively and not only qualitatively. Surely, at this point we cannot fully answer this question. What we will do, instead, is to describe schematically the general features that an objective collapse model must have in order to successfully describe the black hole formation and evaporation process. Of course, these ideas are of a generic nature; studies involving concrete toy-model examples (in 2 dimensions) have been recently considered in \cite{SUDet, SUDet2}. It is also worth mentioning that various researchers believe that any non-unitary modification of quantum theory introduces insurmountable problems. However, a careful analysis of the issue indicates that such beliefs are mistaken (see \cite{UW}).
 
Our starting point will be a collapse theory which is both relativistic and applicable to fields. An example of such a theory is presented in \cite{Bed:11} where the quantum state of a system is connected to a Cauchy hypersurface $\Sigma$ and the evolution from one such hypersurface to another hypersurface $\Sigma'$ is given by 
 \begin{equation} \label{Tomonaga}
 d_x \Psi (\phi; \Sigma') = \left\lbrace -i J(x) A(x) d\omega_x - (1/2) \lambda^2 N^2(x)d\omega_x 
 +\lambda N(x) d W_x
 \right\rbrace\Psi(\phi; \Sigma) ,
 \end{equation}
where $ d\omega_x $ is the infinitesimal space-time volume separating $\Sigma$ and $\Sigma'$, $\lambda $ is the CSL coupling constant, $ J(x)$ is an operator constructed out of matter fields, $W_x$ is a Brownian motion field and $A(x)$ and $N(x)$ are operators that modify the state of an auxiliary quantum field (for more details we remit the reader to \cite{Bed:11}).

Next, we look for modifications of such a model capable of describing correctly the transition from an initially pure state into a mixed one, or, more precisely, capable of accounting for the evolution of the quantum state of the matter field from $ {\mathcal{I} }^-$ to ${\mathcal{I} }^+$. In other words, we need to ensure that the model leads to the enormous amount of information loss and entropy creation that characterize the transition between the state of the matter fields before and after the black hole creation and annihilation. In order to achieve this, we need to secure the fact that the non-unitary behaviour becomes large in the present context but remains small in ordinary situations (because we know that the deviations from quantum theory should be very small in regimes where the theory has been tested).\footnote{For instance, there are strong experimental bounds on the parameter $\lambda$ appearing in the GRW and CSL theories (see \cite{Bas:13} and references therein).} We propose, then, that the overwhelming part of the non-unitary and information non-conserving dynamics takes place in regions close to the singularity (see Figure 2), or, more precisely, in regions near to what will replace the singularity (remember that we are assuming that quantum gravity resolves it). 

What we are proposing might be achieved by replacing the parameter $\lambda$ in Eq. (\ref{Tomonaga}) by some function that characterizes the curvature of the gravitational environment. For instance, $\lambda$ might be replaced by a function of some geometrical scalar such as the Weyl scalar. Such a choice would guarantee an enhancement of non-unitary behaviour near black holes. Moreover, results of phenomenological studies of theories such as GRW or CSL strongly suggest that the $\lambda$ parameter cannot be universal but has to depend on the mass of the particles in question (with more massive particles associated with more ``intense'' collapse rates, see \cite{Pea.Squ:94}). And, of course, a coupling that depends on the space-time curvature, such as the one we are proposing, is a very natural way to implement such dependence. Furthermore, the choice of the Weyl scalar to substitute $\lambda$ seems also to be in line with Penrose's ideas regarding the connection between the Weyl tensor and entropy (see for instance arguments regarding the {\it ``Weyl curvature hypothesis''} in \cite{Pen:04}). Certainly, the parameter that is chosen in order to substitute $\lambda$, must be such that leads to a generation of entropy that matches the standard estimates of entropy generation in black hole formation and evaporation. This is of course a non-trivial requirement, and it could be the case that no function of the Weyl scalar does the job. At any rate, comparison of the predictions of the model we are proposing with standard estimates can be used to constrain some features of the theory. Either way, in the next section we draw from the ideas presented above to propose a link between spontaneous collapse events and microscopic virtual black holes.
\section{Black holes and dynamical reduction}
\label{B}
Collapse events in dynamical reduction theories represent the fundamental source of non-unitarity, and thus, of information loss. It may seem natural, then, to think of them in analogy with black holes, which, as we saw, also break unitarity and lose information. This analogy leads us to a provocative idea. According to the path integral formulation of quantum mechanics, all possible trajectories of a given system contribute to its quantum characterization. Now, if the system in question is some sort of ``quantum space-time,'' provided by some (yet to exist) satisfactory quantum gravity theory, its possible trajectories will inevitable involve arbitrarily small black holes which will form in connection with appropriately localized density fluctuations. And, because of Hawking's radiation (assuming some level of continuity   between large and very small black holes), such black holes are expected to rapidly evaporate. But we saw already that such evaporation leads to localized sources of information loss. The conclusion, then, is that all quantum processes should display some level of breakdown of unitarity and information loss. Or, in other words, that information loss and breakdown of unitarity must be seen as a feature present in all physical situations. 

All of the above leads us to propose the following: collapse events can be regraded as microscopic and virtual versions of black hole formation and evaporation processes. Or, in other words, that it is not only that collapses and black holes exhibit the same behaviour but that they are in fact the same thing. Of course, when actually attempting to construct a theory where this is the case, the fundamental collapse events must somehow be postulated {\it ab initio} as part of the axioms of the theory. The idea, however, is that at the end one must develop a self-consistent picture where, at the appropriate limit, the correct effective descriptions must be recovered. So, for example, when dealing with relatively large black holes, one must be able to recover the semiclassical treatment. However, when considering black holes on smaller scales, the resulting behaviour must be describable in terms of fundamental collapse events.
\section{Conclusions}
\label{C}
We have reviewed the essential aspects of the ongoing debate concerning the fate of information in the evaporation of black holes {\it via} Hawking radiation. Such situation is one in which the tension between quantum theory and gravitation takes a vary dramatic form, so much that it has lead part of the community to describe the process as a paradox, and to propose dramatic ways to avoid the unpleasant conclusions. At the same time, another part of the community has argued, equally vehemently, that the scenario does not at all lead to a paradox. We hope to have contributed to clarify the nature of those disagreements. But not only that, we have also argued in favour of a possible resolution of the problem that has not been consider in much detail before.\footnote{Two notable exceptions are \cite{Pen:04} and \cite{Haw:00}. Penrose argues for a proposal based on statistical considerations in the context of a box containing an evaporating black hole in equilibrium with its environment. Hawking, in turn, points out that information loss in macroscopic black holes implies information loss in microscopic, virtual black holes, from which he concludes that quantum evolution cannot be unitary (as is well known, he latter changed his mind on the subject).}
 
Before wrapping up, it is worth commenting on a relatively new aspect of the discussion that has emerged from the so-called AdS/CFT conjecture. According to such proposal, certain theories involving gravitation in an asymptotically anti-de Sitter $n$-dimensional space-time $M$ are completely equivalent to some conformal field theories without gravity on the $(n-1)$-dimensional boundary of $M$, $\partial M$. Accordingly, the process of formation and evaporation of a black hole in $M$ must correspond to some process in $\partial M$ that, given that does not involve black holes or gravity, must possess a unitary, information preserving evolution. As a result, many researchers in this field have been driven to conclude that at least one of the following facts must be incorrect:
\begin{enumerate}
\item The ADS/CFT conjecture.
\item Black hole evaporation destroys information.
\end{enumerate}
We will not discuss here the evidence and arguments in support of 1 (the literature on the subject is large; see \cite{ADS1, ADS2} for a recent review). The point the we do want to stress is that the strong conviction of its validity has lead people to explore scenarios which abandon central tenets of contemporary physics, often forcing rather unsettling conclusions (like the violations of the equivalence principle proposed in \cite{Fire:12} or the suggestion in \cite{Abhay} of strong deviations from general relativity, due to quantum effects, in regimes one would not naturally expect them).
 
The point, however, is that the disjunctive between 1 and 2 above is predicted on the assumption that quantum evolution is always unitary. If, on the contrary, the Schrödinger equation is replaced with a modification involving stochastic and non-unitary aspects, the contradiction among principles is removed and replaced by a much milder quantitative puzzle: can the extent of the modification be such that, on the one hand, the success of standard quantum theory at the experimental level is preserved, while, at the other, its effects at the CFT side of the duality (the part not involving black holes) are compatible with the magnitude of information loss and non-unitarity that one deduces for the asymptotic observers on the gravitational side (that involving the black hole evaporation)? The approach we are proposing seems to offer a path that retains the validity of 1 and 2, while at the same time, avoids drastic moves. 
 
As far as we know, at this point there are no completely satisfactory relativistic objective collapse models, much less one capable of incorporating gravity in a consistent way. Still, we think that the view we have presented here might not only contain the insights that could lead us to a full resolution of the issue at hand, but also to the construction of a consistent and self-contained quantum theory of gravity. 
\bibliographystyle{ieeetr}
\bibliography{biblioOCC}

\begin{thebibliography}{10}

\bibitem{Haw:75}
S.~Hawking, ``Particle creation by black holes,'' {\em Commun. Math. Phys.},
  vol.~43, pp.~199--220, 1975.

\bibitem{CCC}
R.~Wald, ``Gravitational collapse and cosmic censorship,'' 1997.
\newblock {g}rqc/9710068.

\bibitem{Fire:12}
A.~Almheiri, D.~Marolf, J.~Polchinski, and J.~Sully, ``Black holes:
  Complementarity or firewalls?,'' 2012.
\newblock {a}rXiv:1207.3123.

\bibitem{Br1}
S.~L. Braunstein, ``Black hole entropy as entropy of entanglement or it's
  curtains for the equivalence principle,'' 2009.
\newblock {a}rXiv:0907.1190v1 [quant-ph].

\bibitem{Br2}
S.~L. Braunstein, S.~Pirandola, and K.~Zyczkowski, ``Better late than never:
  Information retrieval from black holes,'' {\em Phys. Rev. Lett.}, vol.~110,
  no.~101301, p.~5 pages, 2013.

\bibitem{SEP}
G.~Ghirardi, ``Collapse theories,'' in {\em The Stanford Encyclopedia of
  Philosophy} (E.~N. Zalta, ed.), winter 2011.

\bibitem{Pea:07}
P.~Pearle, ``How stands collapse {I},'' {\em J. Phys. A: Math. Theor.},
  vol.~40, p.~3189–3204, 2007.

\bibitem{SUDet}
S.~K. Modak, L.~Ortíz, I.~Peña, and D.~Sudarsky, ``Black holes: Information
  loss but no paradox.'' arXiv:1406.4898, 2014.

\bibitem{SUDet2}
S.~K. Modak, L.~Ortíz, I.~Peña, and D.~Sudarsky, ``Non-paradoxical loss of
  information in black hole evaporation.'' arXiv:1408.3062 [gr-qc], 2014.

\bibitem{UW}
W.~G. Unruh and R.~M. Wald, ``On evolution laws taking pure states tomixed
  states in quantum field theory,'' {\em Phys.Rev. D}, vol.~52, pp.~2176--2182,
  1995.

\bibitem{Bed:11}
D.~Bedingham, ``Relativistic state reduction model,'' {\em J. Phys.: Conf.
  Ser.}, vol.~306, pp.~1--7, 2011.

\bibitem{Bas:13}
A.~Bassi, K.~Lochan, S.~Satin, T.~Singh, and H.~Ulbricht, ``Models of
  wave-function collapse, underlying theories, and experimental tests,'' {\em
  Rev. Mod. Phys.}, vol.~85, p.~471, 2013.

\bibitem{Pea.Squ:94}
P.~Pearle and E.~Squires, ``Bound-state excitation, nucleaon decay experiments,
  and models of wave-function collapse,'' {\em Phys. Rev. Lett.}, vol.~73,
  pp.~1--5, 1994.

\bibitem{Pen:04}
R.~Penrose, {\em The Road to Reality}.
\newblock Knopf, 2004.

\bibitem{Haw:00}
S.~Hawking, ``Quantum black holes,'' in {\em The Nature of Space and Time}
  (S.~Hawking and R.~Penrose, eds.), pp.~37--60, Princeton University Press,
  2000.

\bibitem{ADS1}
O.~Lunin and S.~D. Mathur, ``Ads/cft duality and the black hole information
  paradox,'' {\em Nucl.Phys. B}, vol.~623, p.~342–394, 2002.

\bibitem{ADS2}
L.~Fidkowski, V.~Hubeny, M.~Kleban, and S.~Shenker, ``The black hole
  singularity in ads/cft.,'' {\em JHEP}, vol.~0402, no.~014, 2004.

\bibitem{Abhay}
A.~Ashtekar, V.~Taveras, and M.~Varadarajan, ``Information is not lost in the
  evaporation of 2-dimensional black holes,'' {\em Phys.Rev.Lett.}, vol.~100,
  no.~211302, 2008.

\end{thebibliography}
\end{document}